\documentstyle[preprint,aps,floats]{revtex}

\tighten

\begin{document}

\preprint{\vbox{\hbox{JHU--TIPAC--96007} \hbox{hep-ph/9603389}}}

\title{A New Interpretation of the Observed Heavy Baryons}

\author{Adam F. Falk}

\address{Department of Physics and Astronomy, The Johns Hopkins
University\\ 3400 North Charles Street, Baltimore, Maryland 21218}

\date{March 1996}

\maketitle

\begin{abstract}

I suggest that the conventional assignment of quantum
numbers to the observed charm and bottom baryons is not correct,
as these assignments imply large violation of the heavy spin-flavor and 
light
$SU(3)$ symmetries.  I propose an alternative interpretation of the 
observed
states, in which the symmetries are preserved.  If these novel assignments
are right, there is a new state with mass approximately 2380~MeV which
decays to $\Lambda_c+\gamma$, and another with mass approximately
5760~MeV which decays to $\Lambda_b+\gamma$.  Although such states
have not been seen, neither are they excluded by current
analyses.
\end{abstract}


\vfill\eject

The past few years have seen the discovery of many new hadrons containing a
single charm or bottom quark.  Such states fall
into representations of heavy quark spin-flavor $SU(4)$ and light flavor
$SU(3)$ symmetries, up to heavy quark corrections of order
$\Lambda_{\rm QCD}/2m_Q$ and $SU(3)$ corrections of order 
$m_q/\Lambda_\chi$.
Enough have now been discovered to
make possible detailed tests of the relations implied by the symmetries.
In the heavy meson sector, these predictions are known to work well
for the ground states and the lowest $P$-wave excitations~\cite{FaMe95}.
Not only the spectroscopy, but the widths and even the decay angular
distributions are consistent with a simultaneous heavy quark and chiral
$SU(3)$ expansion.  Hence one is tempted to hope that the symmetry
predictions for heavy baryons are also well satisfied. However, in 
contrast to
the mesons,
for the baryons there are certain symmetry relations which appear to be 
badly
violated, although others appear to work well.

While it is possible that the symmetry breaking corrections
are just larger than expected, such an explanation would
offer no insight into why some relations behave better than others.  In
this letter, I will propose that the problem is instead that the 
conventional
assignment of quantum numbers to the observed charm and bottom baryons is 
not
correct. I will show how one can satisfy all the symmetry relations at the
expected level by assigning
new quantum numbers to the known resonances.  An exciting consequence is 
the
existence of additional light excitations which only decay radiatively.  
Such
states are not presently ruled out, and this prediction presents a well 
defined
and
conclusive test of the proposal.

I begin with a review of baryon spectroscopy in the heavy quark limit,
$m_c,m_b\to\infty$.  In this limit, heavy quark pair production and
chromomagnetic interactions are suppressed, so the angular momentum and
flavor quantum numbers of the light degrees of freedom become good
quantum numbers.  I will refer to these light degrees of freedom as a
``diquark''; in doing so, I assume nothing about their properties other 
than
that they carry certain spin and flavor quantum numbers.  For simplicity, I
will also restrict myself for the moment to heavy charm baryons, since the
enumeration of states for bottom baryons is precisely analogous.

In the quark model, the lightest diquark has isospin $I=0$, total spin
$s_\ell=0$ and orbital angular momentum $L_\ell=0$.  With diquark 
spin-parity
$J_\ell^P=0^+$, this leads to the heavy baryon $\Lambda_c$,
with total $J^P=\case12^+$.  The strange analogue of the
$\Lambda_c$ is the $\Xi_c$, with $I=\case12$.  Because of Fermi statistics,
there is no doubly strange state with $s_\ell=0$.
There is a nearby excitation of the $\Lambda_c$, in which the diquark is 
in the
same
orbital state, but with $I=s_\ell=1$.  This leads
to a doublet of heavy baryons consisting of the $\Sigma_c$, with
$J^P=\case12^+$, and the $\Sigma^*_c$, with $J^P=\case32^+$.  As with all 
heavy
doublets, the chromomagnetic hyperfine splitting between these states is of
order $\Lambda_{\rm QCD}^2/m_c$.  The strange analogues of the
$\Sigma_c$ and $\Sigma^*_c$ are respectively the $\Xi'_c$ and $\Xi^*_c$, 
and
there are also the doubly strange states $\Omega_c$ and $\Omega^*_c$.

The diquark may be excited further by adding a unit of orbital angular
momentum,
$L_\ell=1$.  More precisely, this is true in the constituent quark model, 
which
guides our intuition that resonances with these
quantum numbers might be close by.  When $I=s_\ell=0$, the excited diquark 
has
total spin-parity $J_\ell^P=1^-$, and the heavy baryon states are the
$\Lambda_c^*(\case12)$ and the $\Lambda_c^*(\case32)$.  When $I=s_\ell=1$, 
one
finds diquarks with $J_\ell^P=0^-$, $1^-$ and $2^-$, leading to the odd 
parity
heavy baryons $\Sigma^*_{c0}$, $\Sigma^*_{c1}(\case12,\case32)$ and
$\Sigma^*_{c2}(\case32,\case52)$.  There are also excited $\Xi_c$ and
$\Omega_c$ baryons.  The spectroscopy of the charm baryons is summarized in
Table~\ref{hqstates}, along with the allowed decays of the states.  Two
channels are listed where there is the possibility that either is 
kinematically
dominant.
\begin{table}
  \begin{tabular}{llllllrl}
  Name&$J^P$&$s_\ell$&$L_\ell$&$J^P_\ell$&$I$&$S$&Decay\\
  \hline\hline
  $\Lambda_c$&$\case12^+$&0&0&$0^+$&0&0&weak\\
  $\Sigma_c$&$\case12^+$&1&0&$1^+$&1&0&$\Lambda_c\gamma$, $\Lambda_c\pi$\\
  $\Sigma^*_c$&$\case32^+$&1&0&$1^+$&1&0&$\Lambda_c\pi$\\
  $\Xi_c$&$\case12^+$&0&0&$0^+$&$\case12$&$-1$&weak\\
  $\Xi'_c$&$\case12^+$&1&0&$1^+$&$\case12$&$-1$&$\Xi_c\gamma$, $\Xi_c\pi$\\
  $\Xi^*_c$&$\case32^+$&1&0&$1^+$&$\case12$&$-1$&$\Xi_c\pi$\\
  $\Omega_c$&$\case12^+$&1&0&$1^+$&0&$-2$&weak\\
  $\Omega^*_c$&$\case32^+$&1&0&$1^+$&0&$-2$&$\Omega_c\gamma$\\
  $\Lambda_c^*(\case12)$&$\case12^-$&0&1&$1^-$&0&0&
     $\Sigma_c\pi$, $\Lambda_c\pi\pi$\\
  $\Lambda_c^*(\case32)$&$\case32^-$&0&1&$1^-$&0&0&
     $\Sigma^*_c\pi$, $\Lambda_c\pi\pi$\\
  $\Sigma^*_{c0}$&$\case12^-$&1&1&$0^-$&1&0&$\Lambda_c\pi$\\
  $\Sigma^*_{c1}(\case12,\case32)$&$\case12^-$, $\case32^-$
     &1&1&$1^-$&1&0&$\Lambda_c\pi$\\
  $\Sigma^*_{c2}(\case32,\case52)$&$\case32^-$, $\case52^-$
     &1&1&$2^-$&1&0&$\Lambda_c\pi$\\
  \end{tabular}
  \caption{Charm baryon states in the heavy quark limit.  Here $s_\ell$,
$L_\ell$ and $J^P_\ell$ refer respectively to the spin, orbital angular
momentum, and total spin-parity of the light diquark, while $I$ is isospin 
and
$S$ strangeness.  The given decay channel is the one which is expected to 
be
dominant, if kinematically allowed.  The enumeration of the bottom baryon
states is analogous.}
  \label{hqstates}
\end{table}

The masses of these states satisfy a number of heavy quark and
$SU(3)$ symmetry relations.  There are three independent constraints which
relate the bottom and charm systems,\begin{mathletters}\label{allhqrels}
\begin{eqnarray}
   \Lambda_b-\Lambda_c &=&\overline B-
   \overline D=3340\,{\rm MeV}\,,\label{hqrel1}\\
   \overline\Sigma_b-\Lambda_b &=&
   \overline\Sigma_c-\Lambda_c\,,\label{hqrel2}\\
   {\Sigma^*_b-\Sigma_b\over\Sigma^*_c-\Sigma_c} &=&
   {B^*-B\over D^*-D}=0.33\,,\label{hqrel3}
\end{eqnarray}
\end{mathletters}%
where in (\ref{hqrel1}) and (\ref{hqrel3}) I have inserted the isospin 
averaged
heavy
meson masses~\cite{PDG}.  Here the states stand for their masses, and a bar
over a state denotes
the spin average over the heavy multiplet of which it is a part.  This spin
average, which cancels the hyperfine interaction between the heavy quark 
and
the collective light degrees of freedom, takes the form $(D+3D^*)/4$ for 
the
ground state heavy mesons and $(\Sigma_c+2\Sigma_c^*)/3$, {\it etc.}, for 
the
spin-$(\case12,\case32)$ heavy baryon doublets.  The hyperfine
relation~(\ref{hqrel3}) is more commonly written in terms of the ratio
$m_c/m_b$, to which each side is equal, but I prefer a form in which the 
quark
masses are not introduced explicitly.  The corrections to~(\ref{hqrel1})
and~(\ref{hqrel2}) are expected to be of order
$\Lambda_{\rm QCD}^2(1/2m_c-1/2m_b)\sim50\,$MeV.  The corrections to
(\ref{hqrel3}) could be at the level of~25\%.

The light flavor $SU(3)$ relations are trivial in the exact symmetry limit,
where, for example, $\Sigma_c=\Xi'_c=\Omega_c$.  In this form, they
are also badly violated.  If one includes the corrections linear in
$m_s$, one finds four independent ``equal spacing rules'' for states 
within the
charm
(or bottom) system~\cite{Savage},
\begin{mathletters}\label{allsu3rels}
\begin{eqnarray}
   \Omega_c-\Xi'_c &=& \Xi'_c-\Sigma_c\,,\label{su3rel1}\\
   \Omega^*_c-\Xi^*_c &=& \Xi^*_c-\Sigma^*_c\,,\label{su3rel2}\\
   \Sigma^*_c-\Sigma_c &=& \Xi^*_c-\Xi'_c
   = \Omega^*_c-\Omega_c\,,\label{su3rel3}\\
   \overline\Sigma_c-\Lambda_c &=&
   \overline\Xi\vphantom{\Xi}^*_c-\Xi_c\,.\label{su3rel4}
\end{eqnarray}
\end{mathletters}%
Here I neglect isospin violation and electromagnetic effects. The
chiral corrections to the  relations~(\ref{su3rel1})--(\ref{su3rel3})
are expected to be small~\cite{Savage}.  The relation~(\ref{su3rel4})
is not on the same footing as the others, since it
relates states in two {\it different\/} $SU(3)$ multiplets. It is actually 
a
combined $SU(3)$ and heavy quark symmetry relation.  The leading
corrections to it are, in principle, of order $m_s$, and cannot be
calculated.  However, one's intuition from the quark model is that this
relation should
be reasonably well satisfied, and indeed the counterparts in the charmed 
meson
sector, such
as $D_{s1}-D_s=D_1-D$, work to within 10~MeV.  In fact all of the heavy 
quark
and $SU(3)$ relations for the charm and bottom mesons work
beautifully~\cite{FaMe95}.

So far, a dozen charm and bottom baryon states have been discovered. I list
them, along with their masses and observed decays, in Table~\ref{baryons}.
However, the names conventionally given to the strongly decaying states 
imply
certain assumptions
about their quantum numbers and properties.  Since it is precisely these
assumptions which I want to challenge, I instead identify the observed
resonances
by the modified names listed in the first column of Table~\ref{baryons}. 
For
simplicity, I have averaged over
isospin multiplets, since isospin breaking is small and not at issue here.
\begin{table}
  \begin{tabular}{llllll}
  State&Mass (MeV)&Ref.&Decay Channel&Conventional&
    Proposed\\
  \hline\hline
  $\Lambda_c$&$2285\pm1$&\cite{PDG}&weak&$\Lambda_c$&$\Lambda_c$\\
  &(2380)&&$\Lambda_c+\gamma$&absent&$\Sigma_c$\\
  
$\Sigma_{c1}$&$2453\pm1$&\cite{PDG}&$\Lambda_c+\pi$&$\Sigma_c$&$\Sigma^*_c$
\\
  $\Sigma_{c2}$&$2530\pm5\pm5$&\cite{SKAT}&$\Lambda_c+\pi$&
    $\Sigma^*_c$&$\Sigma^*_{c0}$\ (?)\\
  $\Xi_c$&$2468\pm2$&\cite{PDG}&weak&$\Xi_c$&$\Xi_c$\\
  $\Xi_{c1}$&$2563\pm15$\ (?)&\cite{WA89}\tablenote{The mass of the 
$\Xi_{c1}$
is estimated from the plots presented by WA89.  Only
one of the two isospin states has been observed.}
    &$\Xi_c+\gamma$&$\Xi'_c$&$\Xi'_c$\\
  $\Xi_{c2}$&$2644\pm2$&\cite{CLEO95}&$\Xi_c+\pi$&$\Xi^*_c$&$\Xi^*_c$\\
  $\Omega_c$&$2700\pm3$&\cite{E687}&weak&$\Omega_c$&$\Omega_c$\\
  $\Lambda^*_{c1}$&$2593\pm1$&\cite{PDG,CLEO94}
    &$\Sigma_{c1}+\pi\to\Lambda_c+2\pi$&$\Lambda_c^*(\case12)$&
    $\Lambda_c^*(\case32)$\\
  $\Lambda^*_{c2}$&$2627\pm1$&\cite{CLEO94}&$\Lambda_c+\pi+\pi$&
    $\Lambda_c^*(\case32)$&$\Lambda_c^*(\case12)$\\
  \hline
  
$\Lambda_b$&$5623\pm5\pm4$&\cite{PDG,CDF96}&weak&$\Lambda_b$&$\Lambda_b$\\
  &(5760)&&$\Lambda_b+\gamma$&absent&$\Sigma_b$\\
  $\Sigma_{b1}$&$5796\pm3\pm5$&\cite{DELPHI}&$\Lambda_b+\pi$&
    $\Sigma_b$&$\Sigma^*_b$\\
  $\Sigma_{b2}$&$5852\pm3\pm5$&\cite{DELPHI}&$\Lambda_b+\pi$&
    $\Sigma^*_b$&$\Sigma^*_{b0}$\ (?)\\
  \end{tabular}
  \caption{The observed heavy baryon states, with their conventional and
proposed identities.  Isospin multiplets have been averaged over.  
Experimental
errors ($\pm {\rm stat.}\pm {\rm sys.})$ are included where
significant; where they are small, statistical and systematic errors have, 
for
simplicity, been added in quadrature.  The approximate
masses of the proposed new states are given in parentheses.}
   \label{baryons}
\end{table}

The conventional identities of the observed heavy
baryons are given in the fourth column of Table~\ref{baryons}.  How well 
do the
predictions of heavy quark and $SU(3)$ symmetry fare?  The heavy quark
constraints~(\ref{hqrel1}) and~(\ref{hqrel2}) are both satisfied to within
$10\,$MeV.  However, the hyperfine relation (\ref{hqrel3}) is in
serious trouble.  One finds
$(\Sigma_b^*-\Sigma_b)/(\Sigma_c^*-\Sigma_c)\approx0.73\pm0.13$, too
large by a factor of two!  To be conservative, I have ignored the 
correlation
between the errors on the $\Sigma_b$ and the $\Sigma^*_b$, hence
overestimating the total uncertainty.  It is clear that  {\it to take these
data seriously is to identify a crisis for the application of heavy quark
symmetry to the charm and bottom baryons.}

Neither is the situation perfect for the $SU(3)$ relations.  The first 
equal
spacing rule (\ref{su3rel1}), with the well measured masses of the
$\Sigma_c$ and the $\Omega_c$, yields the prediction $\Xi'_c=2577\,$MeV,
somewhat large but probably within the experimental error.  The second rule
(\ref{su3rel2}) cannot be tested, as the $\Omega^*_c$ state has not yet 
been
found.  Inserting the measured $\Sigma_c$, $\Sigma^*_c$ and $\Xi_c^*$ 
masses,
the
third rule~(\ref{su3rel3}) may be rearranged to yield the prediction
$\Xi'_c=2567\,$MeV,
reasonably consistent both with (\ref{su3rel1}) and with experiment.  
However,
the final
$SU(3)$ relation~(\ref{su3rel4}) fails by approximately 80~MeV, an order of
magnitude worse than for the charmed mesons!  Such an enormous discrepancy 
is
quite surprising and disappointing.

What are we to make of this situation, in which one heavy quark and one
$SU(3)$ relation fail so badly?  Given that there is no reason to doubt the
quoted experimental errors, perhaps we must simply accept that there are
large corrections, that somehow these important symmetries are 
inapplicable to
heavy baryons.  However, with their striking success in the heavy meson
sector, {\it especially for spectroscopy\/}, it is tempting to look for a 
new
point of
view from which the symmetry predictions are better behaved.

In this light, I propose to reinterpret the experimental data under the
constraint
that the heavy quark and $SU(3)$ symmetries be imposed explicitly.  Then 
if we
identify, once again, the observed $\Xi_{c1}$ with the
$\Xi'_c$ state, the $SU(3)$ relations~(\ref{allsu3rels}) lead to the novel 
mass
prediction $\Sigma_c\approx2380\,$MeV!  If so, the $\Sigma_c$ cannot be
identified with the observed $\Sigma_{c1}$; in fact, it can be identified 
with
no resonance yet to have been reported.  However, since at this mass the
$\Sigma_c$ can decay only radiatively, $\Sigma_c\to\Lambda_c+\gamma$, it
is quite possible that it exists but so far has been overlooked.

The observed $\Sigma_{c1}$ is now identified as the $\Sigma^*_c$.  In the
bottom baryons, there is a similar reassignment:  the $\Sigma_b$ is now
assumed to be below $\Lambda_b+\pi$ threshold and to decay radiatively,
while the $\Sigma_{b1}$ is identified as the $\Sigma^*_b$.  As for the
observed $\Sigma_{c2}$ and $\Sigma_{b2}$, they are possibly $I=1$,
$L_\ell=1$ excitations, such as the $\Sigma^*_{c(0,1,2)}$.  While one might
naively estimate that the masses of these states should be larger than 
those
of the $\Lambda^*_c(\case12)$ and $\Lambda^*_c(\case32)$, a substantial
spin-orbit coupling
could lower the mass of the state $\Sigma^*_{c0}$ by of order $200\,$MeV.
Hence I tentatively identify the observed $\Sigma_{c2}$ and $\Sigma_{b2}$
respectively as the $\Sigma^*_{c0}$ and $\Sigma^*_{b0}$.

The poorly behaved symmetry relations improve dramatically in this
scenario.  For example, let us take the masses of the new states to be
$\Sigma_c=2380\,$MeV and $\Sigma_b=5760\,$MeV.  Then the hyperfine
splitting ratio (\ref{hqrel3}) improves to
$(\Sigma_b^*-\Sigma_b)/(\Sigma_c^*-\Sigma_c)=0.49$, and the $SU(3)$
relation (\ref{su3rel4}) between the $s_\ell=0$ and $s_\ell=1$ states is
satisfied
to within $5\,$MeV.  The heavy quark relation~(\ref{hqrel1}) is unaffected,
while the constraint~(\ref{hqrel2}) for the
$\overline\Sigma_Q$ excitation energy is satisfied to within $20\,$MeV, 
which
is quite reasonable.  Only the
$SU(3)$ equal spacing rules~(\ref{su3rel1}) and~(\ref{su3rel3}) suffer 
mildly
from the
change.  Taken, as before, as a prediction for the mass of the $\Xi_c'$, 
the
former relation now fails by
$23\,$MeV.  The latter now fails by $8\,$MeV, but the discrepancies are in
{\it opposite\/} directions, and the two relations cannot be satisfied
simultaneously by shifting the mass of the $\Xi'_c$.  With these new
assignments, intrinsic
$SU(3)$ violating corrections of the order of $15\,$MeV seem to be 
unavoidable.

With respect to the symmetry predictions as a whole, the new scenario is an
enormous improvement over the old.  The heavy quark and $SU(3)$ flavor
symmetries have been resurrected.  We can improve the agreement further if 
we
allow the
measured masses to vary within their reported $1\sigma$ errors.  One set of
allowed
masses is
$\Sigma_c=2375\,$MeV, $\Sigma^*_c=2453\,$MeV, $\Xi'_c = 2553\,$MeV,
$\Xi^*_c=2644\,$MeV, $\Sigma_b=5760\,$MeV, and $\Sigma^*_b=5790\,$MeV.  For
this
choice, the $SU(3)$ relations (\ref{su3rel1}) and (\ref{su3rel3}) (taken as
predictions for the $\Xi'_c$ mass) and (\ref{su3rel4}) are satisfied to 
within
$15\,$MeV, $13\,$MeV and $4\,$MeV, respectively.  The hyperfine ratio
(\ref{hqrel3}) is $(\Sigma_b^*-\Sigma_b)/(\Sigma_c^*-\Sigma_c)=0.38$, and
$\overline\Sigma_b-\Lambda_b$ is equal to
$\overline\Sigma_c-\Lambda_c$ to within $15\,$MeV.  This is better
agreement with the symmetries than we even have a right to
expect.

As appealing a scenario as this is, certain problems do remain.  First,
while the radiatively decaying states $\Sigma_c$ and $\Sigma_b$ have not 
been
ruled
out, neither have they yet been identified.  In the end, their discovery 
or the
absence
thereof will be the defining test of this proposal.  Second, the excited 
baryon
$\Lambda^*_{c1}$ is seen to decay via the two step process
$\Lambda^*_{c1}\to\Sigma_{c1}+\pi\to\Lambda_c+2\pi$,
while the two pion decay of the $\Lambda^*_{c2}$ is 
nonresonant~\cite{CLEO95}.
If the
observed states $\Lambda^*_{c1}$ and $\Lambda^*_{c2}$ are identified with 
the
heavy doublet $\Lambda^*_c(\case12)$ and $\Lambda^*_c(\case32)$, then the 
first
stage in the
decay of the $\Lambda^*_{c1}$ is dominated by $S$-wave pion
emission~\cite{Cho}.  If so,
the spin of the $\Lambda^*_{c1}$ is the same as that of the $\Sigma_{c2}$,
namely
$J=\case32$.  Hence the excited $I=0$ doublet must be inverted, with
$\Lambda^*_c(\case32)<\Lambda^*_c(\case12)$.  Perhaps this situation is
somewhat
unnatural, perhaps not.

However, the least satisfactory feature of this scenario is the 
identification
of
the $\Sigma_{b2}$ as the $\Sigma^*_{b0}$ state, with $s_\ell=L_\ell=1$ and
$J^P_\ell=0^-$.  The DELPHI analysis~\cite{DELPHI} of the masses, 
production
and decay properties of the $\Sigma_{b1}$ and $\Sigma_{b2}$ explains in an
elegant
and nontrivial manner the surprisingly low observed polarization of
$\Lambda_b$'s produced at the $Z^0$.~\cite{ALEPH,FaPe94}.  The analysis was
predicated, of course, on the
conventional assignment of quantum numbers; now this nice explanation of
$\Lambda_b$ depolarization is lost.  Worse, while the $S$-wave decay
$\Sigma^*_{b0}\to\Lambda_b+\pi$ must be isotropic, there appears to be a 
large
anisotropy
in the direction of the pion in 
$\Sigma_{b2}\to\Lambda_b+\pi$~\cite{DELPHI}.
The reported
deviation from an isotropic distribution is about $2.5\sigma$.  If this 
result
is confirmed, the observed $\Sigma_{b2}$ state must be something else, 
such as
a radial excitation of the $\Sigma_b^*$.

Finally, it is worth noting that nonrelativistic consituent quark models 
(see,
for example, the many papers cited in Ref.~\cite{Savage}) typically do not
favor such a light $\Sigma_c$ and $\Sigma_c^*$ as I have suggested here.  
In
fact, such models cannot be reconciled simultaneously with the heavy quark
limit and with the reported masses of the $\Sigma_b$ and $\Sigma_b^*$.  
Hence,
the predictions of this letter follow experiment in pointing to physics 
beyond
the constituent quark model.  While the historical usefulness of this 
model for
hadron spectroscopy may well lead one to be suspicious of the $\Sigma_b$ 
and
$\Sigma_b^*$ data, such speculation is beyond the scope of this 
discussion.  I
have taken the masses and errors of all states as they have been reported 
to
date; as they evolve in the future, so, of course, will the theoretical
analysis.

While such issues are important, the smoking gun here is the
prediction of new heavy baryon excitations of approximately $100\,$MeV,
decaying radiatively to $\Lambda_c$ and $\Lambda_b$.  If confirmed, this 
will
be the
most unexpected and striking prediction yet to be obtained from heavy quark
symmetry.  If not, and if the reported data are correct, we will have to 
accept
the failure of heavy spin-flavor
and light $SU(3)$ symmetry to describe the charm and bottom baryon states.

It is a pleasure to thank Jon Bagger, Mike Booth, Bob Fletcher, Mike Luke 
and
Tom Mehen for helpful conversations, and John Yelton and Don Fujino for
invaluable correspondence concerning the CLEO experiment.  This work was
supported by the National Science Foundation under Grant
No.~PHY-9404057 and National Young Investigator Award No.~PHY-9457916, by 
the
Department of Energy under Outstanding Junior Investigator Award
No.~DE-FG02-94ER40869, and by the Alfred P.~Sloan Foundation.

\end{document}